%% file: 2017_0033.tex
\documentclass[referee]{raa}

\usepackage{graphicx,times,ulem,titletoc,rotating,color}             %for PS/EPS graphics inclusion, new

\begin{document}

\title{Sky Subtraction for LAMOST}

\volnopage{Vol.0 (200x) No.0, 000--000}      %%preserved for Editor. DOn't remove!
\setcounter{page}{1}          %%starting page, preserved for Editor. DOn't remove!

\author{Zhong-Rui Bai
      \inst{1,2}
   \and Hao-Tong Zhang
      \inst{2}
   \and Hai-Long Yuan
      \inst{2}
   \and Guang-Wei Li
      \inst{2}
   \and Jian-Jun Chen
      \inst{2}
   \and Ya-Juan Lei
      \inst{2}
   \and Hui-Qin Yang
      \inst{2}
   \and Yi-Qiao Dong
      \inst{2}
   \and Gang Wang
       \inst{2}
   \and Yong-Heng Zhao
      \inst{1,2}
}

\institute{University of Chinese Academy of Sciences, Beijing 100049, China\\
        \and
             Key Lab for Optical Astronomy,
             National Astronomical Observatories, Chinese Academy of Sciences,
             Beijing 100012, China; {\it htzhang@bao.ac.cn}\\
}

\date{Received~~2009 month day; accepted~~2009~~month day}

\abstract{Sky subtraction is the key technique in data reduction of multi-fiber spectra.
Knowledge of the related instrument character is necessary to determine the method adopted in sky subtraction.
In this study, we described the sky subtraction method designed for
  LAMOST(Large sky Area Multi-Object fiber Spectroscopic Telescope) survey.
The method has been intergrated into LAMOST 2D Pipeline v2.6 and applied to data of LAMOST DR3 and later.
For LAMOST, sky emission line  calibration  is used to alleviate
 the position-dependent (thus time-dependent) $\sim4\%$ fiber throughput uncertainty and the small wavelength instability (0.1\AA ) during observation.
PCA (Principal Component Analysis) sky subtraction further reduces $25\%$ of the sky line residual of the OH lines in the red part of the LAMOST spectra after the
  mater sky spectrum, which is derived from a B-spline fit of 20 sky fibers in each spectrograph,  is adjusted by sky emission line and subtracted from each fiber.
Further analysis shows that our wavelength calibration accuracy is about 4.5km/s, and the average sky subtraction residuals are about
  3\% for sky emission lines and  3\% for continuum region.
The relative sky subtraction residuals
 vary with the moon light background brightness, could reach as low as 1.5\% for the sky emission line regions in the dark night.
Tests on the F stars of  both similar sky emission line strength and similar object continuum intensity
 show that the sky emission line residual of LAMOST is smaller than those of SDSS survey.
}

\authorrunning{Z.-R. Bai, H.-T. Zhang, H.-L. Yuan, et al. } %author_head in even pages
\titlerunning{LAMOST sky subtraction}  % title_head in odd pages

\maketitle

%{\tableofcontents}

\section {Introduction}

Multi-object spectroscopy with optical fibers, which is  a leap-type development for astronomical observation due to its ability of
 simultaneously observing much more objects  than the traditional long slit spectroscopy, has been routinely carried out over three decades.
Unlike slit or multi-slit systems,
the sky spectrum can not be sampled closely adjacent to the object both on the focal plane and on the CCD  in multi-object fiber spectroscopy.
This difference makes both the observation and the data reduction strategy of multi-fiber spectroscopy differs from the slit spectroscopy.
The standard procedure of sky subtraction for multi-object fiber spectroscopy involves  using a subset of fibers (sky fibers)
to measure the sky background simultaneously with  the object fibers (\cite{Wyse92}; \cite{Watson98})
A master sky spectrum is constructed from the sky fibers and then subtracted from each object+sky spectrum.

In practice, the sky subtraction accuracy is considered good in the range 1\%-2\% (\cite{Elston89}; \cite{Cuby94}).
The limitation comes from various reasons, including focal-ratio degradation of the fibers, internal scattered light, variation of the sky,
telecentricity effects (\cite{Wynne93}), cross talk from adjacent fibers and poor determination of the fiber transmittance
(\cite{Elston89}; \cite{Watson98})

A number of astronomers have explored techniques to improve the sky subtraction.
Observational strategies such as beam-switching (\cite{barden93}; \cite{Puech14}; \cite{Rodrigues12})
and nod-and-shuffle (N+S, \cite{GBK01}; \cite{Sharp10})  could help to eliminate the throughput difference between
fibers and obtain higher sky subtraction accuracy, but extra cost of exposure time or CCD space is inevitable.
For the standard observation mode, strong night sky emission lines are often used to calibrate the relative transmission
 of fibers to an  accuracy of sky subtraction  better than 2\%(e.g. \cite{Lissandrini94}).
 Principal component analysis (PCA) is another  well-established  technique that
has been applied in the sky subtraction for fiber spectroscopic in the recent 10 years
after its  first demonstration by \cite{Kurtz00}.
\cite{Wild05} presented a technique to remove the residual OH features based on the PCA of the residual of the sky subtracted sky spectra in the SDSS DR2 and achieved a dramatic improvement in the quality of a large fraction of SDSS spectra,
particularly for the fainter objects such as the high-redshift quasars.
\cite{Sharp10} demonstrated that the PCA is more efficient than the
N+S technique for observations in the sky limited regime with durations of 10-100 h.
\cite{Soto16} introduced ZAP, an approach to sky subtraction based on PCA,
which is likely to be of a  useful tool to substantially improve  the sky subtraction accuracy.

In this paper we describe the sky subtraction technique for the Guo Shou Jing Telescope
(a.k.a. LAMOST, \cite{cui12}).
The technique has been intergrated into LAMOST 2-dimensional (2D) Pipeline v2.6 and applied to LAMOST Data Release 3 (DR3) and later.
The telescope and  instrument character involved in sky subtraction are introduced in Sec. 2. The sky subtraction methodology for LAMOST is present in Sec. 3.
 The sky subtraction accuracy is analysed in Sec. 4. Some discussions and conclusion are  in Sec. 5 and  6, respectively.

\section {LAMOST}

\subsection {LAMOST Instruments and Observation}
LAMOST is a special Schmidt telescope which allows
both a large aperture (effective aperture of 3.6m-4.9m) and a wide field of view (FoV, hereafter) of 5$^{\circ}$ (\cite{cui12}).
 4000 optical fibers are accommodated on the focal plane, each of which is 320 microns in diameter,
equivalent to 3.3 arcseconds in the sky.
Each fiber is driven by a fiber positioning unit containing two stepping motors,
by which all fibers can be positioned  simultaneously in less than 10 minutes.
The fibers are grouped into 16  spectrographs,
in each of which the light beam is split into blue arm (370-590nm) and
red arm (570-900nm) by a dichroic mirror  then  registered by a 4k$\times$4k CCD camera in each arm.
The spectral resolution is about 2.5\AA\ in blue arm and 4\AA\ in red arm.

To optimize the observing efficiency and mitigate the fiber cross talk, targets in LAMOST survey are grouped into bright, medium and faint plan according to their $r$-band magnitude.
The on-site astronomer decides which one to execute based on the moon phase and the weather condition.
Fainter plans are always observed in darker nights with better weather conditions.
 Multiple exposures, usually three, are taken to obtain enough Signal to Noise Ratio(S/N) and to remove cosmic rays.
The typical exposure time  of one sub-exposure for bright, medium and faint plans are 600, 900 and 1800 seconds, respectively.
Twilight flats  are taken at zenith both in the evening and morning for correcting the instrument difference between fibers and three Mercury-Cadmium-Neon-Argon arc-lamp frames are taken at the beginning, end and the middle of the observational night, respectively.

\begin{figure}[bp]
\centering
\includegraphics[width=0.95\textwidth]{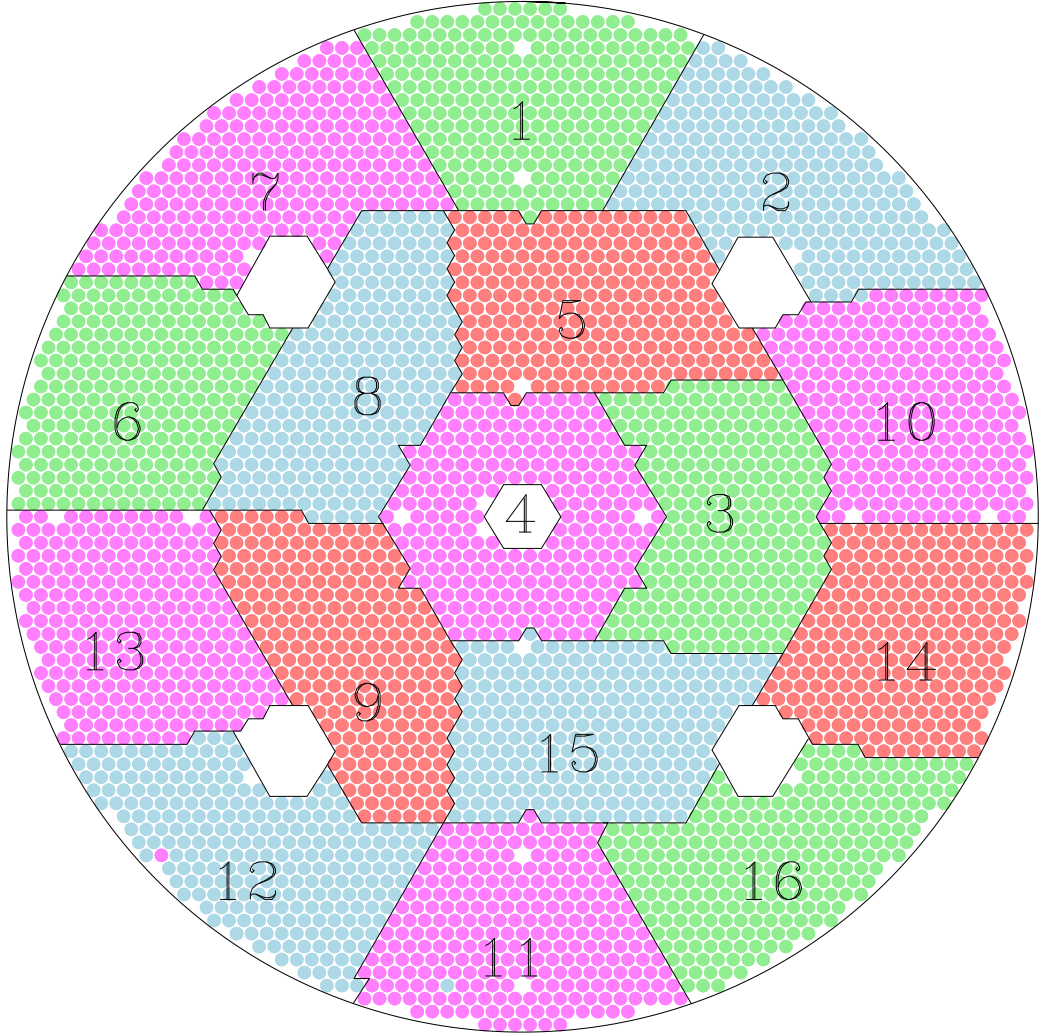}
\caption{LAMOST fiber division scheme. The fibers on the focal plane   are divided to 16 regions, each feeds into one spectrograph.
The colored circles indicate position of the fibers and the solid line is the border of the spectrographs.
The ID of the spectrographs are  marked in the corresponding region.} \label{lamostfp}
\end{figure}

The 5$^{\circ}$ diameter FoV, or 20 square degrees, are divided to 16 pieces, each piece(250 fibers, 1.25 square degrees) are feeded into one spectrograph,
as shown in Fig.\ref{lamostfp}.
  In dark nights, the night sky shows stable gradients on scales of degrees (\cite{Wyse92}), so the spatial variation of sky inside one
spectrograph is insignificant. 20 of 250 fibers, distributed homogeneously both on the sky and on the CCD, are dedicated to sample the sky spectra.
The traditional sky subtract method is performed spectrograph by spectrograph.
The master sky spectrum is constructed from those sky fibers, as shown in the following sections.

\subsection {Dark Night Sky Spectrum at LAMOST Site}
\begin{figure}[bp]
\centering
\includegraphics[width=0.95\textwidth]{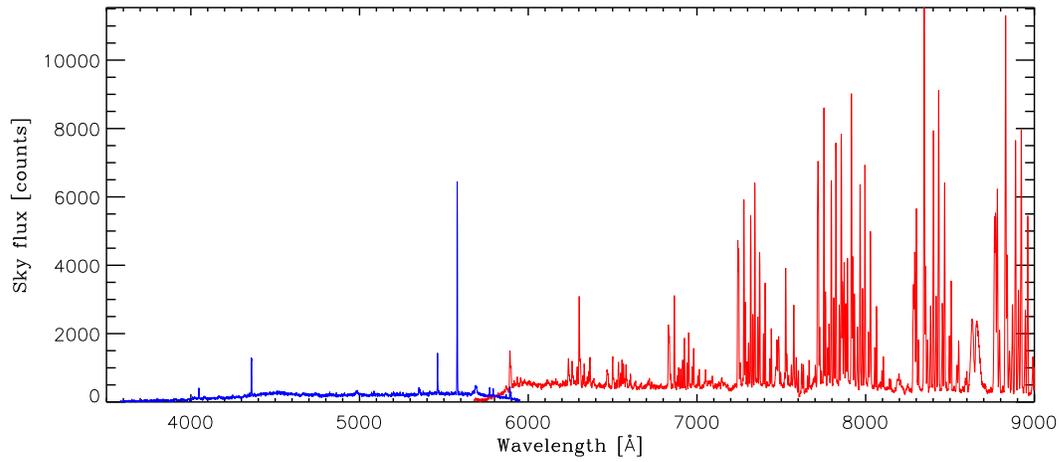}
\caption{Moonless night sky spectra taken by LAMOST on  September 12th, 2015.
Spectra of the blue and the red arm are indicated by the color.
The x-axis is in original CCD pixels but wavelength calibrated.} \label{skysample}
\end{figure}

A typical dark night sky spectrum observed by LAMOST is shown in Fig.\ref{skysample}.
Except a few emission lines
 come from the artificial light pollution, most of the distinctive features of the night-sky spectrum,
including the continuum, absorption lines and most of the emission lines, are due to natural processes.

The continuum of the night-sky are contributed by zodiacal light, the starlight, the extragalactic light,
and the reflected solar light (\cite{Benn98}).
The airglow, emitting from various processes of atoms and molecules in the upper atmosphere,
 produces the [OI]5577\AA, 6300\AA, 6363\AA\ lines, the O band at 8600-8700\AA,
 NaD 5890-5896\AA, as well as the OH bands in the red and IR, known as the Meinel bands (\cite{Meinel50}).
The light pollution mostly comes from the street lamps including mercury lamps and sodium lamps.
 For LAMOST, the mercury streetlight produce strong narrow lines at 3651\AA, 3663\AA, 4047\AA, 4358\AA, 5461\AA, 5770\AA\ and 5791\AA;
and the sodium  lamp contributes to NaD lines while other Na lines are relatively weak.

\subsection {LAMOST Instrumental Effect}
\subsubsection{Instrument difference \label{fibertrans}}
To subtract sky spectrum with sky fibers,  accurate calibration of the  relative instrument difference from the  mirror of the telescope to the CCD pixel is very important. Generally, those differences include the vignetting effect of the telescope; the throughput difference between fibers which may  be caused either by the intrinsical difference due to various reasons(e.g. fiber length, polishing of the end face, etc) or by the misalignments between the fiber and the optical axis of input light beam (\cite{Wyse92}); the vignetting effect of the spectrograph and the pixel to pixel difference of the CCD.
 Usually the correction is achieved by using a uniformly illuminated  flat field, either the twilight sky or a screen at the telescope pupil. But for LAMOST,
 a special designed reflecting Schmidt telescope with very large field of view, the difference can not be corrected directly by either kind of flat field.

As shown in \cite{Xue08}, since the aperture of LAMOST changes with the telescope pointing, the vignetting effect
is as large as 30\% across the LAMOST field  or 10\% for the spectrographs at the edge of the LAMOST field, depending on the target declination and hour angle.
Considering  the fact that twilight only considered to be homogenous within degrees around the
 zenith and  the very limited observation time during twilight, no twilight flat can compensate this position  dependent effect.
 It is as well impossible to  build  a large dome flat screen at the telescope
 pupil, which is the 4.5-meter Schmidt reflector,  and illuminate it uniformly at the same pointing as observation.
  So  the twilight flat is only taken to
 correct the instrument response along the wavelength direction and the instrument difference that is relatively stable with time and position.

\begin{figure}[bp]
\centering
\includegraphics[width=0.5\textwidth]{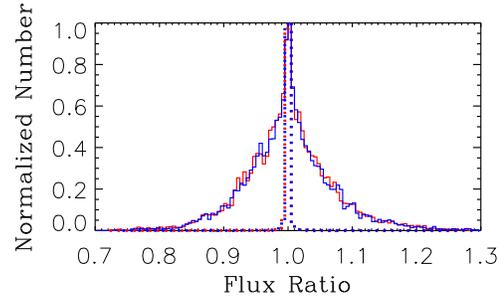}
\caption{Flux ratio of the dome flat field.
The solid lines show the flat field flux ratio of different fiber position, while the dot-dashed lines are  of the same fiber position.
The blue and the red lines are from the red and the blue arm,respectively.
The $\sigma$ of the Gaussian fitting of the solid lines are 4.77\% and 4.83\% for the blue and the red, respectively,
while those for the dot-dashed lines are 0.280\% and 0.283\%.
} \label{fluxratio_fp}
\end{figure}

 On the other hand, twisting, bending  and stress  on the fiber will change the focal ratio degeneration of the light beam at the fiber output end,
 leading to the change of the fiber throughput. Unfortunately, LAMOST suffers such effect when the fiber positioner  put the fiber  to a new position.
 Fig.\ref{fluxratio_fp} shows  the  result of a test of the  fiber throughput changing with fiber positions  on   January 19th, 2013.
 During the test, the telescope pointing and the focal plane position was fixed;  A dome flat screen illuminated by incandescent
  lamps was put in front of the focal plane;
 Dome flat field exposures were taken  for two sets of fiber position in turn, so that neighboring exposures are of different fiber position and every other
 exposures are of the same fiber position; Totally 14 exposures  were taken for each fiber position, respectively.
 Since the position difference of  each fiber between neighbouring exposures are relatively small, the flat field brightness difference between the two position of the
same fiber cloud be ignored,
 the only reason that cause the  difference  between neighbouring exposures of the same fiber should be the stress put on the fiber by the fiber positioner.
 The statistic of the flux ratio between each pair of the neighbouring exposures and the flux ratio between the exposure of the same fiber position shows that
 the throughput uncertainty caused by  the fiber positioner is about 4.8\%, much larger than the uncertainty of the poisson noise(0.2\%), which will leave large  sky subtraction residuals  if not corrected properly.

As shown above, the fiber-to-fiber throughput difference depends on the telescope pointing and the position of the fiber,  could not be corrected
by a twilight flat field. Currently, the only possible solution is calibrating with the strong sky emission lines  that go through the same light path as the target, such as  [OI]5577\AA \ in the blue and  some of  OH lines in the red.

\subsubsection {Image shift}
 Variations of circumstance temperature and gravity (if moving with the telescope) will induce instability of the spectrograph,
 thus the image shifts in both the spatial and dispersion directions, which will lead to trace and wavelength calibration error, then finally
 the bad sky subtraction. For LAMOST, the image shift is mainly caused by the temperature variation and the consuming loss of liquid nitrogen which
 put weight directly on the CCD camera. As shown in Table\ref{t_waveshift}, most of the shift in wavelength direction during the whole night is smaller than 0.1\AA,
 while there are certain spectrographs(e.g. spectrograph 16)  that are larger. Also could be seen from the table is that the shift is not homogenous during the night.
 Except taking arc image between each exposure,  using the sky emission line is a cheaper but robust solution to calibrate the shift.

\section {Methodology}

Sky subtraction is one of the final steps in the LAMOST 2D data reduction pipeline, but
 dependent on the quality of the previous steps.
The spectra are extracted from the raw science data using the fiber trace obtained from flat field frame.
The initial wavelength solution is obtained by arc-lamp frame
and the initial fiber-to-fiber transmittances are estimated by the twilight flat field spectra.
The sky emission lines are used to fine-tune the wavelength solution and the fiber-to-fiber transmittances.
After that, the master sky spectra are created from the sky sampling fibers, then subtracted from the object spectra.
The object spectra are flux-calibrated, different exposures are co-added and interpolated  to a logarithmically-spaced wavelength scale, $\Delta\lg\lambda=10^{-4}$.
Finally, PCA sky subtraction is performed on the co-added spectra in the wavelength range of 7200-9000\AA, where most sky emission lines lie.

This paper will focus on the sky subtraction, skipping other steps like
flux extraction, arc-lamp wavelength calibration, flat-fielding, flux calibration and spectra co-addition, which will be described in detail
in a forthcoming paper (Bai et al.,  in preparation).
We start from the extracted flux, assuming the initial wavelength solution and initial fiber transmittances has been performed.

\subsection {Sky Emission Lines Identification}

The sky emission lines could be easily identified after the initial wavelength calibration with arc lamp.
In the blue arm, the sky emission lines are relatively sparse and the street light lines are too weak to offer  reliable calibration,
 so only the strong  airglow line [OI]5577\AA\ are used.
While in the red arm, there are  bunches of strong emission lines, such as OH bands.
It is not easy to identify single lines in this region, yet after a careful comparison between the observed spectra
and  literature(\cite{Osterbrock96}; \cite{Osterbrock97}), 13 single lines and 6 doublets are selected, as listed in Table\ref{shiftline}. For the doublet,
 the intensity of the two lines are similar and the separation of the doublet is less than  0.5\AA, so that they can be treated as single lines under
 LAMOST resolution($\sim4$\AA), then only the  average wavelength of the two lines are adopted  in the table.

For the $i$th selected line with wavelength  $\lambda_i$\ in an individual fiber,
the profile is fitted with a S\'ersic function and a linear background within $\pm8$\AA\ around the line center $\lambda_i$:

\begin{equation}\label{eq:1}
f(\lambda)=\alpha{e}^{-\frac{|\lambda-{\lambda}_{i}|^\delta}{\delta\gamma^\delta}}+a\lambda+b,
\end{equation}
where $f(\lambda)$ is the flux corrected by the twilight flat, $\alpha$ and $\delta$ are the parameters of the S\'ersic function and
$a\lambda+b$ is the fitted background continuum.
The intensity of the line is the sum of the continuum subtracted segment:
\begin{equation}\label{eq2}
F_i=\int^{\lambda_i+8}_{\lambda_i-8}{\left[f(\lambda)-a\lambda-b\right]}d\lambda.
\end{equation}
As noticed by \cite{bai17} and \cite{Li15}, some of the LAMOST emission line profile can not be perfectly fitted by a S\'ersic function due to optical aberration and distortion.
So the S\'ersic function is only used to derive the accurate wing of the emission line and the back ground function, not taking part in equation \ref{eq2}.

\subsection {Wavelength Calibration with Sky Lines} \label{waveshift}

 Since the image shift  varies slightly with fibers, the wavelength solution is corrected fiber by fiber. 
The wavelength shift of a sky line is defined as the difference between the literature wavelength $\hat{\lambda}_{i}$ and the initial line center $\lambda_i$:
\begin{equation}
\Delta{\lambda_{i}}=\hat{\lambda}_{i}-\lambda_i,
\end{equation}

Then  $\Delta{\lambda_{i}}$ are fitted with  a linear function of ${\lambda}_{i}$:
\begin{equation}
\Delta{\lambda_{i}}=m{\lambda}_{i}+n,
\end{equation}
 the coefficients $m$ and $n$ are derived by solving the above functions with the least square method.

\begin{figure}[bp]
\centering
\includegraphics[width=0.95\textwidth]{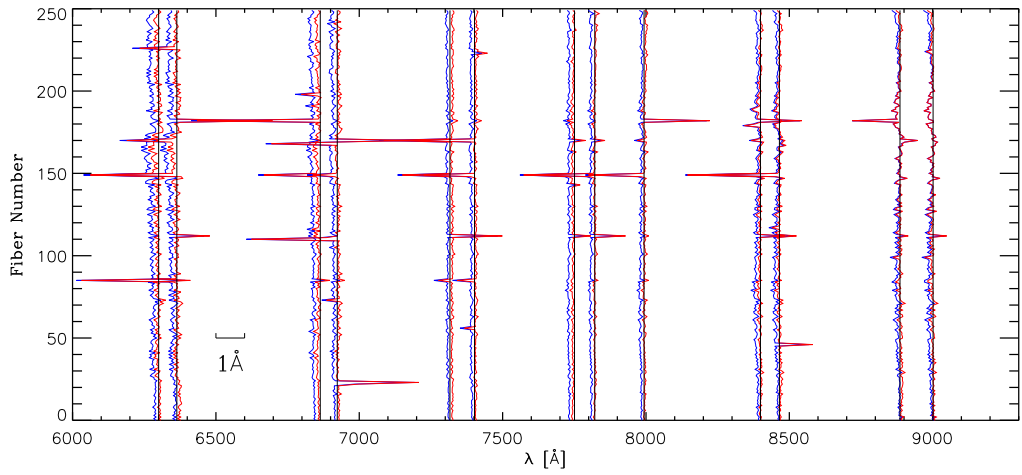}
\caption{An example of the wavelength corrections in a single frame.
The blue and red lines show the measured wavelengths of the sky emission lines based on the arc-lamp solution and the sky emission line fine-tuned wavelength, respectively. While the black lines are
wavelength in the literature.
The distance of the blue and red lines relative to the black are   100 times exaggerated to show the details. The average difference between the initial wavelength solution and
the corrected wavelength is $0.13\pm 0.07$\AA, which is typical for the correction.
The large residuals seen in the plot are caused by bad fibers or bad pixels.} \label{skylineshift}
\end{figure}

Finally, the updated wavelength solution  are obtained by
\begin{equation}
\lambda'={\lambda}+m{\lambda}+n,
\end{equation}
where $\lambda'$ is the updated wavelength.
For blue arm,  $m$ is set to be zero since only [OI]5577\AA\ is used.
An example of the wavelength correction of the red arm are shown in Fig.\ref{skylineshift}.

\subsection {Fiber Transmittance Correction}

As described in Section \ref{fibertrans}, the relative fiber throughput varies with the telescope pointing and the fiber position, this could be  calibrated using the
intensity of the sky emission line $F$, calculated in  Eq.(\ref{eq2}). For the $i$th line in a fiber, the scale factor $s(i)$ could be calculated by:

\begin{equation}
s(i)=\frac{F(i)}{<F(i)>},
\end{equation}
where $F(i)$ is the line flux obtained by Eq.(\ref{eq2}) and $<F(i)>$ is the median  of $F(i)$ over all fibers.

For the blue arm,  the scale factor is the relative scale of $[OI]5577$\AA\  line.
 For the red arm, the median of  $s(i)$ over  19 lines listed in table\ref{shiftline}  is adopted as the final scale factor.
The flat field corrected  spectra is then divided by the scale factor.

\begin{figure}[bp]
\centering
\includegraphics[width=0.95\textwidth]{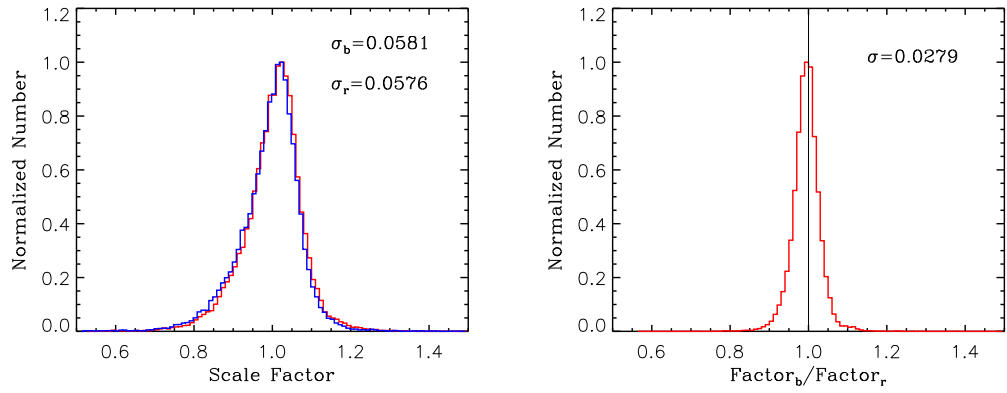}
\caption{Left: Distribution of the scale factors of the blue and the red arms.
Right: Distribution of the ratio of the blue factors to the red one.
The standard deviations are as indicated.} \label{skyscale}
\end{figure}

Fig.\ref{skyscale} shows an example of the distribution of the scale factors on October 26th, 2016.
The scale factor in both the blue and the red arms show similar non-Gaussian distribution with  standard deviation of about 0.058,
while the ratio of the scale factor of the blue arm to the red is close to Gaussian distribution with standard deviation of 0.028.
These indicate that the uncertainty of the fiber throughput induced by the telescope vignetting effect and the fiber positioner
is about 5.8\%, while the  accuracy of scale factor correction is about 2.8\%.

As measured from the data, the uncertainty of the sky emission line intensity is about 1.5\% for [OI] 5577\AA\ and any single OH lines.
Considering $s_r$ is derived from the median of 19 OH lines, the uncertainty of the median is $1.5\%/\sqrt{19}{\approx}0.35\%$ approximately.
From the twilight flat fields observed in two adjacent days, the uncertainty of the twilight sky flat is about 1.6\%.
In  total, the synthetic uncertainty of $s_b/s_r$ is about $\sqrt{1.5^2+0.35^2+2\times{}1.6^2}{\approx}2.7\%$,
consistent with the accuracy of scale factor correction.

\subsection{Master Sky}

\begin{figure}[bp]
\centering
\includegraphics[width=0.95\textwidth]{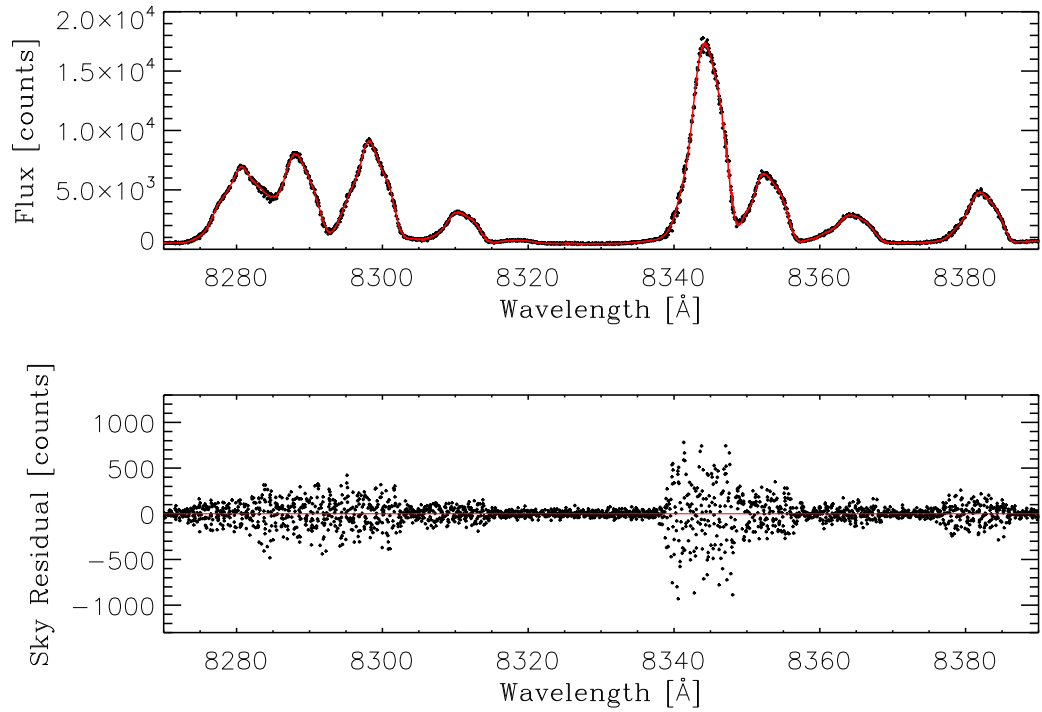}
\caption{An example of sky subtraction for spectra taken on Dec. 31th, 2016.
Upper: Fluxes from 20 sky fibers of one spectrograph corrected by sky emission lines.
The black crosses shows all 20 sky spectra  and the solid red line shows the best fitted master sky.
The discrepancy of different sky fibers is too small to be discriminated in current scale.
Lower: The residuals of the sky subtracted  sky spectra.} \label{bsplinefit}
\end{figure}

After the  wavelength  and fiber transmittance are fine-tuned by the sky emission line,
 a master sky spectrum is created  from the sky fibers in the same spectrograph, using the  B-spline fitting procedure similar to
SDSS 2D pipeline(see SDSS data reduction pipeline $idlspec2d$, \cite{Bolton07}).
Spectra of sky fibers are treated as fluxes in discrete pixels; pixels from  different sky spectra are aligned together  in order of their wavelength.
The master spectrum is fitted in 2 dimensions,  a cubic B-spline function  in the wavelength direction,  allowing the B-spline coefficients
 to vary with the fiber number. The bad pixels are rejected during the fitting.
The  B-spline function is then interpolated back to each fiber, obtaining the final sky spectrum, which will be
subtracted from the object spectrum. Fig. \ref{bsplinefit} shows an example of master sky spectrum
and the sky subtraction residual.

\subsection {PCA Sky Subtraction}

After the master sky is subtracted, each spectrum is flux calibrated, then
different exposures are combined and interpolated to  a logarithmically-spaced wavelength scale,  i.e. $\Delta\log\lambda=10^{-4}$.
PCA is performed on the combined spectra in the range of 7200-9000\AA, where the OH sky emission lines are dominating.
For each spectrograph, about 20 sky subtracted sky spectra are used to generate the components of  PCA.
Both the sky  and object spectra  are first continuum subtracted using  a rolling median filter to remove large-scale structures,
then the eigenvectors and eigenvalues of  PCA are derived from the $\sim$20 sky residual spectra.
For each spectrum, a projection coefficient is calculated for each eigenvector,
and the sum of the 20 most-significant principal eigenvectors weighted by the projection coefficients is
 adopted to derived the sky residual spectrum. This residual is then removed from the  spectrum and the median filtered continuum is added back.
The detail of PCA sky subtraction could be found in \cite{Wild05} and \cite{Sharp10}.

\section{Sky Subtraction Accuracy}
 The sky subtraction routine described here is part of the LAMOST 2D Pipeline v2.6
and has been applied in all LAMOST data later than Data Release 3.
As the sky subtraction are performed in two stages (i.e. the master sky stage and the PCA stage),
 the accuracies of the two stages are analysed separately.

\subsection {Wavelength Calibration Accuracy}

\begin{figure}[bp]
\centering
\includegraphics[width=0.95\textwidth]{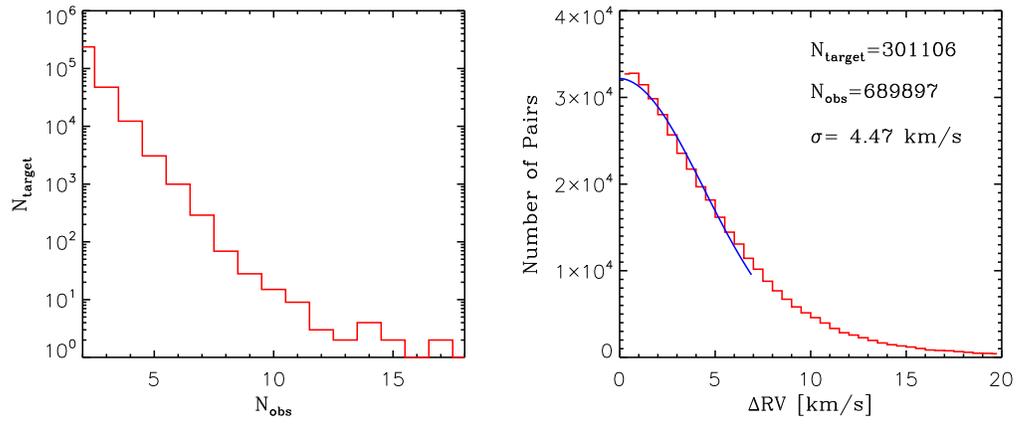}
\caption{Left: The histogram of the repeat observations in LAMOST database, x axis indicate the number of repeat observations.
Right:  Distribution of  radial velocity difference of the repeat observations; the blue curve is the Gaussian fit of the "core" of the distribution.} \label{rvcompare}
\end{figure}

The typical  error of sky line calibration within one observation is about 0.07\AA, as shown in Fig \ref{skylineshift}.
It is not clear whether our calibration is stable between observations, so it is necessary to test it by measuring the  radial velocity(RV) variations of stars.
There are quite a lot stars with multiple observations of more than one day in LAMOST database
and their radial velocities could be used to indicate the stability of our wavelength calibration.
A search of repeated observation of F, G and K stars with S/N over 20 in  LAMOST DR3 results in 689897 spectra  of 301106 stars. 
Every two observations of the same star is defined as a "pair",
so there will be $\mathrm{C}_n^2$ pairs for a star with $n$ observations.
For each pair, $\Delta{}RV$ is defined as $|v_1-v_2|$,
where $v_1$ and $v_2$ are the measured RVs of the two observations, respectively.
The distribution of the RV difference vs the number of observations  is shown in Fig.\ref{rvcompare}.
The standard deviation of the Gaussian fit of the core of the $\Delta{}RV$ distribution is 4.47km/s,
 consistent with the typical wavelength calibration uncertainty (0.07\AA).

\subsection {Sky Subtraction Accuracy of Single Frame}\label{secskyres}

The forthright way to estimate the sky subtraction accuracy is to measure the residual of the sky subtracted sky spectra.
The absolute sky residual is dominated by the shot noise of the original sky flux.
A relative sky residual is defined as the ratio of the absolute sky residual to the sky flux:
\begin{equation}
r_s(\lambda)=\frac{f_{r}(\lambda)}{f(\lambda)}, \label{eqskyres}
\end{equation}
where $f_{r}(\lambda)$ is the absolute residual of the sky spectra and $f(\lambda)$ is the original sky flux.

For sky emission lines, since the typical FWHM of any single sky line is 3-4\AA,
pixels within $\pm${3}\AA\ around the  line center are used to calculate the residuals of the sky emission lines.
In the blue arm,  [OI] 5577\AA\  is measured, while in the red arm  10 strong lines including 7714\AA,
 7750\AA, 7794\AA, 7821\AA, 7853\AA, 7913\AA, 7964\AA, 7993\AA, 8025\AA\ and 8062\AA\  are adopted.

For sky continuum,  the  flux in individual pixel is much lower than that of the emission lines.
To depress the shot noise, instead of using counts of individual pixels in  Eq.(\ref{eqskyres}), the average of
 the continuum in  5470-5560\AA\ and 6000-6200\AA\ region(see Fig.\ref{skysample}) are adopted for the blue and the red arms, respectively:
\begin{equation}
r_s(\lambda)=\frac{\overline{f_{r}(\lambda)}}{\overline{f(\lambda)}}.\label{eq:cave}
\end{equation}

As an example, Fig. \ref{skyresidualsample} shows the distribution of the relative sky residuals on September 20th, 2016.
The dispersion of the residuals, which can be estimated by  $\sigma$ of the Gaussian fitting of the histogram,
is an indicator of the sky subtraction accuracy. As could be seen in Fig.\ref{skyresidualsample},  the histogram of the [OI] 5577 emission line residual could be well
described by a  Gaussian function except at the tail of the histogram, which is mostly caused by  the profile difference
between the master sky spectra and the  optical aberration distorted profile of individual fiber. 
   In the OH line region, the scatter is larger, since not all the 10 lines selected
to measure the residual are from the line list in table \ref{shiftline} to determine the scale factor, also the residual of non single lines will be affected by the neighbouring lines. The residuals in continuum are consistent with the corresponding emission lines, indicating that the scale factor works well in the continuum  region.

\begin{figure}[bp]
\centering
\includegraphics[width=0.95\textwidth]{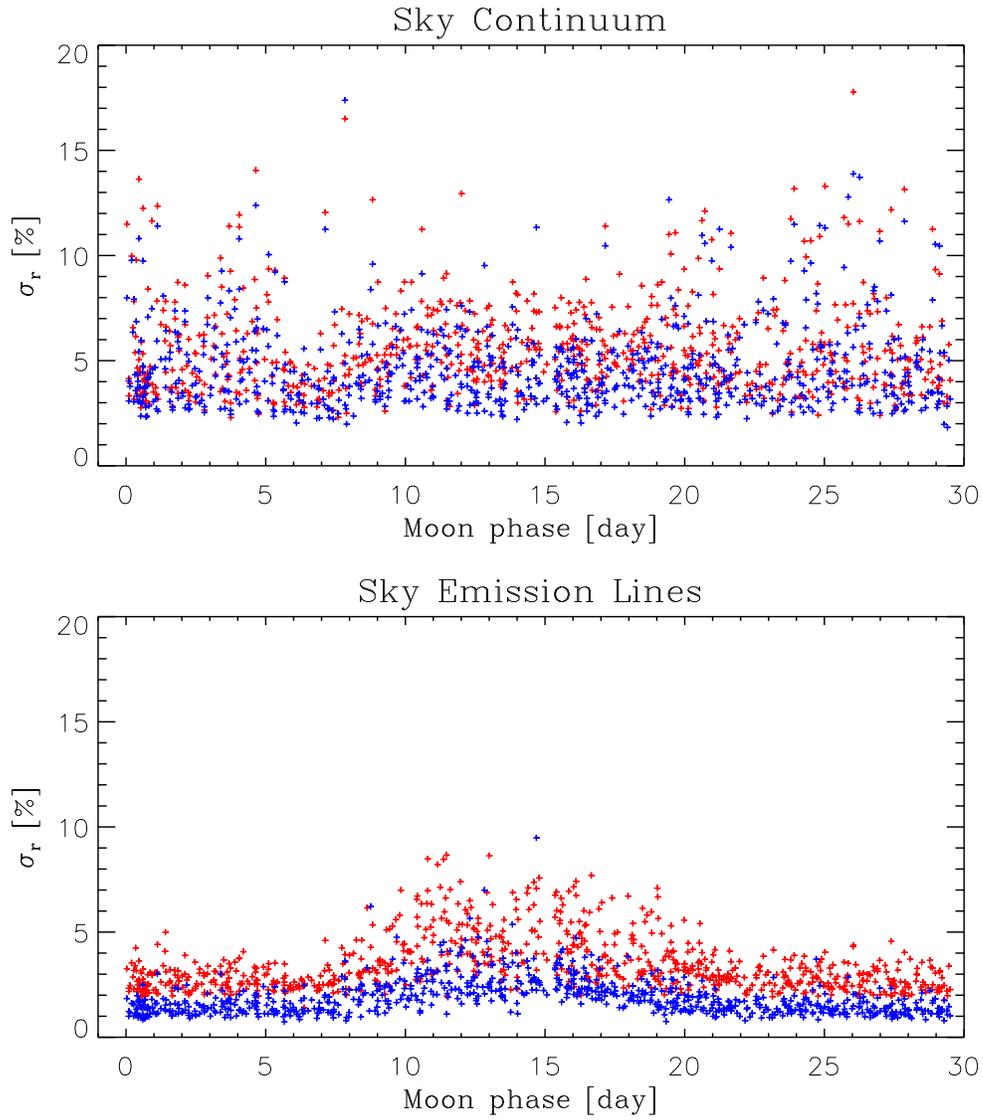}
\caption{Relative sky subtraction residual $\sigma_r$ vs  moon phase.
Upper panel: $\sigma_r$ of the  average sky continuum.
Lower panel: $\sigma_r$ of the  individual pixels of  sky emission lines.
The blue and the red symbols denotes the blue and the red arm, respectively.
The x-axis is the date of moon phase, where 0 denotes new moon and 14.7 denotes full moon.
} \label{skyresidual}
\end{figure}

LAMOST survey is carried out in both bright and dark nights. In moonlit nights, the relative strength of sky emission line to the sky continuum is weaker than that in the darker nights, due to the increase of the background brightness.
To investigate the sky subtraction accuracy dependency on the moon light, we have  traced $\sigma$ for all the LAMOST sky subtracted sky spectra in each night before December 31, 2016 and the results are shown in
Fig.\ref{skyresidual}.

There is no obvious  system tendency of  sky continuum residual $\sigma$ on moon phase. The average values of $\sigma$ for the blue and the red sky continuum residual are comparable to those in Fig.\ref{skyresidualsample}, but with much larger scatter than $\sigma$  of the emission lines. 
There are 3 reasons for the large scatter. The first is that the exposure time varies between
  10 and 30 minutes, then both the continuum and the sky emission line used to calibrate the continuum vary 3 times, thus the relative noise changes  $\sqrt{{3}\times{2}}=2.45$ times. The second is that both the sky continuum and the emission line changes from time to time. The third is that sky continuum does not necessarily vary the same as the sky emission line,  especially in bright nights,  though the difference should be small inside individual spectrograph (see section \ref{scvar}).
  A check of those point with large scatter in dark and grey nights in the upper panel of Fig.\ref{skyresidual} shows
  that those are from observation with short exposures, where both the sky continuum and emission lines are relatively weak.
   The large scatter of continuum residual at the bright nights follow the  similar trend of emission lines in the corresponding moon phase, which is caused by the  short exposure time and the dramatic variation of the  moon light background, thus the increasing of the relative uncertainty of the sky emission lines.

  The relative residuals of the sky emission lines are much smaller in dark nights than the continuum,
with a median level of 1.5\% for [OI] 5577 and 2.2\% for OH lines,
but raising obviously when moon phase is between the 9th and the 21st day,
up to 3\% for [OI] 5577 and 4.5\% for OH lines in full moon nights.

 Considering that  [OI] 5577 emission line is less affected by neighboring lines than the OH lines in the red and
the region used to calculate the scatter in the blue is the region around [OI]5577 itself, it is easy to understand the
smaller residual of the blue arm.

\subsection {Profit of PCA}\label{accmes}
\begin{figure}[bp]
\centering
\includegraphics[width=0.95\textwidth]{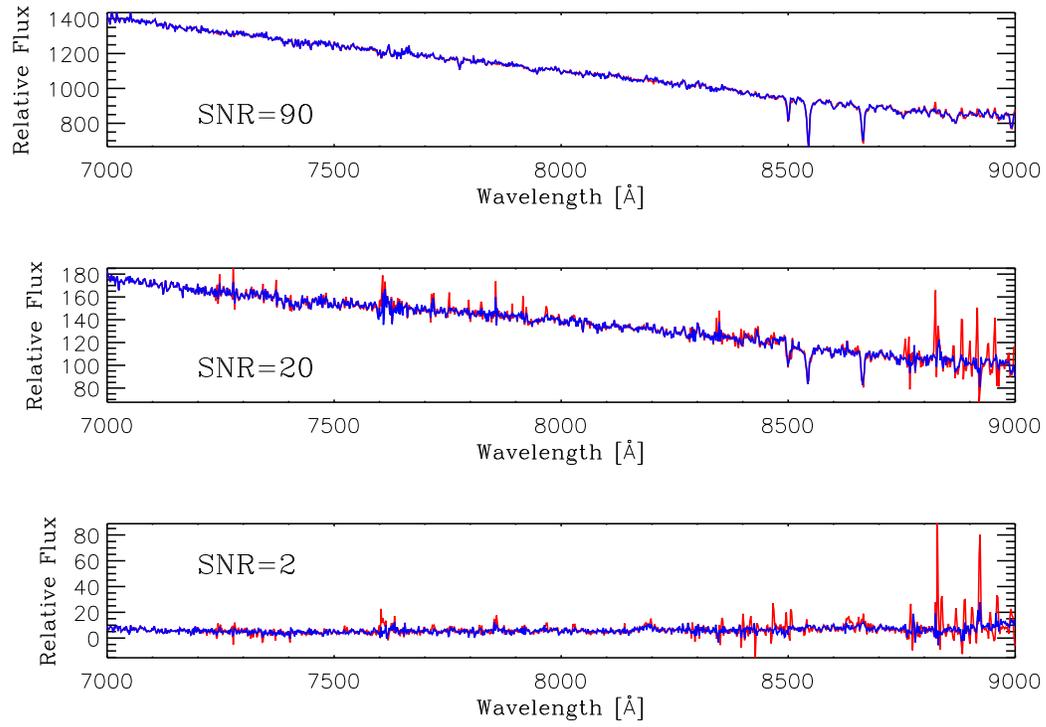}
\caption{Three example spectra with different S/N  in  7200\AA-9000\AA\  band.
In each panel, the master sky  subtracted spectrum is shown in red and the  PCA sky subtracted spectrum is  blue.
} \label{pcasample}
\end{figure}

The PCA sky subtraction focus on the OH emission lines in the range of 7200-9000\AA,
the residual of these lines decreases significantly.
Three examples of different S/N  are shown in Fig. \ref{pcasample}.

\begin{figure}[bp]
\centering
\includegraphics[width=0.5\textwidth]{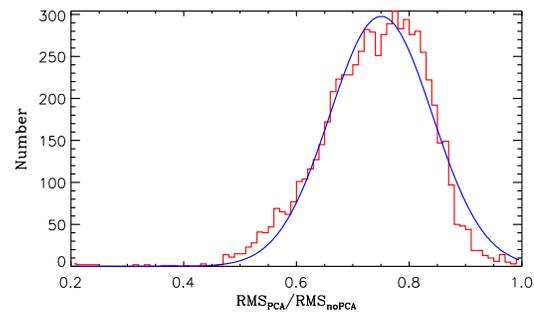}
\caption{The distribution of the ratio of the sky subtraction residual of 7522 F-type stars after the  PCA sky subtraction to that before the PCA sky subtraction.
The red line is the histogram and the blue is the Gaussian fitting of the histogram.} \label{pcacomp}
\end{figure}

To quantify the contribution of PCA subtraction,  the spectra  of 7,522 F-type stars observed on Feb 20, 2016 are analyzed.
For F stars, there are less features in the wavelength 7720-8100\AA, on the contrary, there are abundant sky emission lines in the same region,
so the smoothness of the continuum  could be used to evaluate the quality of the sky subtraction.
The smoothness is defined as:
\begin{equation}
a_t=RMS(\frac{f_{r}-c}{c}), \label{objacc}
\end{equation}
where $f_r$ is the co-added object spectra, $c$ is the pseudo-continuum derived by a rolling median filter
and $RMS$ is  the short for Root Mean Square. Fig.\ref{pcacomp} shows the ratio of
the RMS $a_t$  after the PCA sky subtraction to that before PCA sky subtraction.  All of the stars show smaller RMS after PCA sky subtraction.
The peak occurs at  about 75\%, which indicates that PCA improves the sky subtraction of OH lines about 25\%.
 Incorporating the results of Fig.\ref{skyresidual},
 the median of the final sky subtraction residual of OH lines in the dark and the bright nights could reach as low as 1.7\% and 3.4\%, respectively.

\subsection {Comparison with SDSS}
\begin{figure}[bp]
\centering
\includegraphics[width=0.95\textwidth]{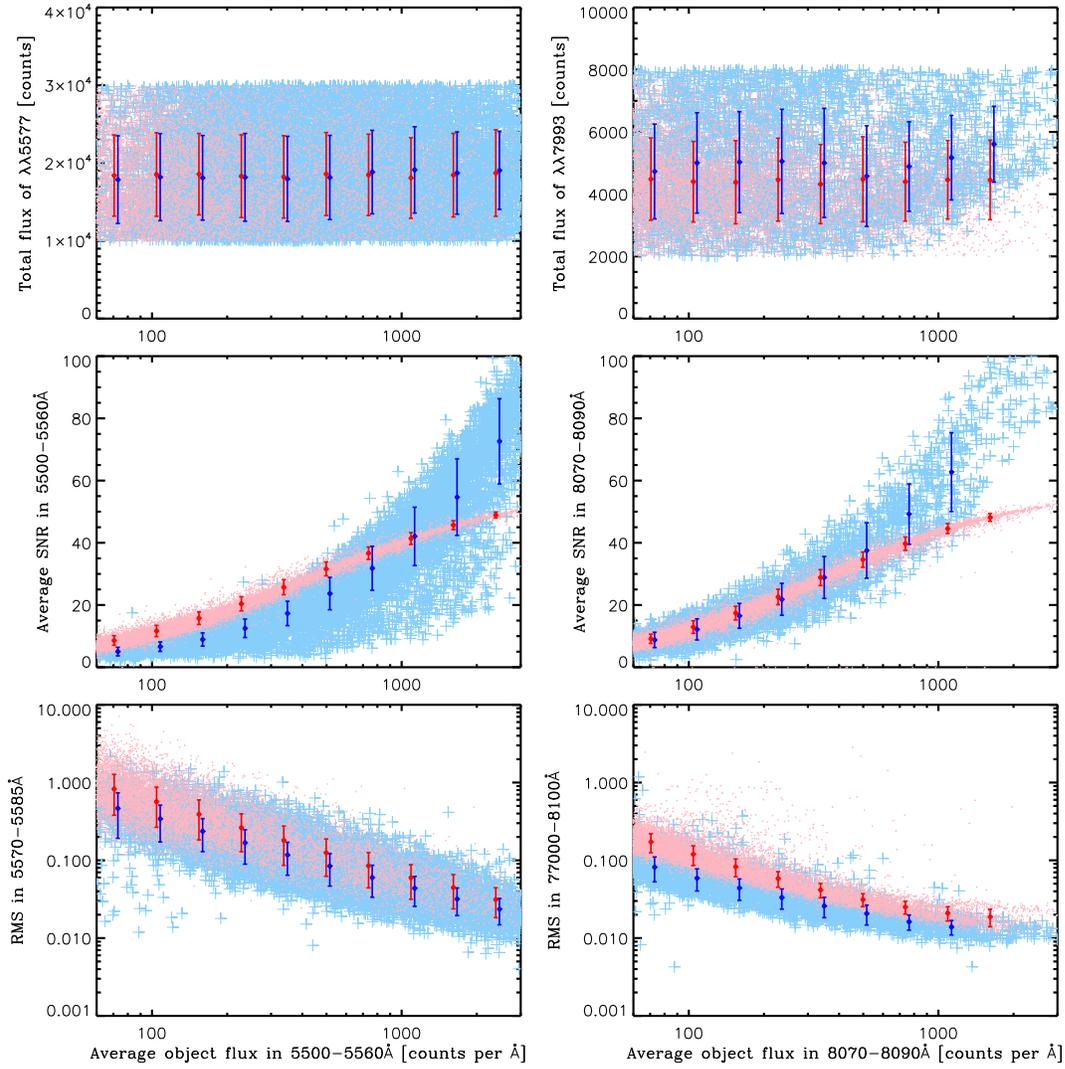}
\caption{
 Comparison of sky emission line subtraction residuals of SDSS and LAMOST F types stars.
The blue symbols denote LAMOST DR3 and the red symbols represent spectra of SDSS DR10 in all panels.
The stars are selected with similar [OI]5577\AA\  integral intensity and similar object continuum flux.
The object flux, indicated in  x axis of all panels,  is calculated as the average counts per angstrom of star continuum in 5500-5560\AA\ and 8070-8090\AA\
for the blue and the red part, respectively.
Top: the star continuum intensity vs the integral flux of sky emission line, [OI]5577\AA\ and OH 7913\AA\ for the blue and the red part, respectively.
Middle: the star continuum intensity vs  S/Ns in the blue and the red part, respectively.
Bottom: the star continuum intensity vs the relative residuals of the sky emission lines, [OI]5577\AA\ and OH lines in 7700-8100\AA\ for the blue and the red part, respectively.
Note that both x and y axes are in log scale, so the error bars are not symmetric.
}
\label{tosdss}
\end{figure}

SDSS (\cite{York00}), similar to LAMOST in resolution and wavelength coverage,
is the most successful low resolution multi-fiber spectroscopic program.
SDSS has higher system efficiency than LAMOST,  18\% and 20\% in blue and red arms, respectively (\cite{Stoughton02}).
Considering the efficiency difference, the signal of the same object in the two surveys are different,
so it is hard to tell which survey shows the smaller  relative sky subtraction residual to the object continuum. Instead of comparing   coincident targets
in both surveys, comparing the targets of both similar strength of sky emission lines and similar counts of object flux is more reasonable.
As shown in the top left panel of Fig.\ref{tosdss}, about 11000   F stars are  selected from both SDSS DR10 and LAMOST DR3
 with similar integral strength of [OI]5577\AA\   and similar object continuum intensity.
Checking the same sample in the red shows that  the OH sky emission lines in LAMOST are much stronger than those in SDSS, so we only keep those LAMOST spectra with the strength of OH 7913\AA\
comparable to those of SDSS, rejecting about 7000 spectra with stronger OH lines, as shown in the top right panel of Fig.\ref{tosdss}.  

When the intensities of the sky emission lines are similar in the blue part of the
two surveys, the sky continuum of LAMOST is brighter than SDSS, thus the S/N of LAMOST stellar continuum in  5500-5560\AA\  is lower than that of SDSS. While in the red, the sky continuum intensity is similar to SDSS, so the
S/N of the two surveys are similar.  As noticed in the middle row of Fig.\ref{tosdss}, the S/N of the two surveys departure
at high object flux region, the reason is as follows.

The S/N of both LAMOST and SDSS spectra are defined as:
\begin{equation}
S/N=flux*\sqrt{invvar}
\end{equation}
where $invvar$ is the inverse variance, which is accumulated during the data processing. In SDSS image processing, it is defined as:
\begin{equation}
invvar=\frac{1}{\sqrt{f+(g*n_{rd})^2+10^{-4}f^2}},
\end{equation}
where $f$ is the object flux counts, $g$ is the gain and $n_{rd}$ is the read-out noise. The 3rd term in denominator is just to limit the S/N below 100.
But in LAMOST image processing, the inverse variance is simply
\begin{equation}
invvar=\frac{1}{\sqrt{f+(g*n_{rd})^2}}.
\end{equation}
The difference of the estimation makes SDSS spectra have lower S/N than LAMOST at the same flux,
especially when the flux is high.

According to Eq.(\ref{objacc}), the residuals in the emission line region of the blue (5570-5585\AA\  around [OI]5577) and the red (7700-8100\AA\ in OH lines region) part are
calculated respectively for each star. The residuals vs their object continuum intensities are plotted in the bottom panel of Fig.\ref{tosdss}.
LAMOST spectra show smaller sky emission line subtraction residuals than SDSS in both the blue and the red part of the spectra, while the red part is more
 significant than the blue. As indicated in the middle left
panel of Fig.\ref{tosdss}, in the blue part, the S/N of the  LAMOST spectra is lower  than SDSS, so the actual effect of LAMOST sky subtraction  should be even
 better than SDSS if they are in the same situation. The two reasons that LAMOST performs  better than SDSS in sky subtraction  may be: 
 LAMOST sky sampling fiber is denser than  SDSS both on the sky (25 fibers per 1.25 square degrees vs  32 fibers per 7 square degrees) and in the CCD image (25 per 250 fibers vs 32 per 640 fibers), so the 
 sky spectra is better represented in LAMOST;   The PCA sky subtraction, which contributes a lot in reducing the sky subtraction residuals in the red part of the LAMOST spectra, is not used for the
 SDSS spectra. Many other details in software and hardware may also explain the different performance of the two survey, but it is hard to make a comprehensive comparison.

\section {Discussion}
The final sky subtraction accuracy is affected by many factors in data  reduction process, most of them could be corrected by the above schemes,
but there are certain problems could not be solved currently, such as the moon light background and the variation of the PSF.

\subsection {Sky Subtraction in Moonlit Night} \label{scvar}
The sky emission line calibration  of the fiber throughput is based on the assumption that the sky background is
homogenous and the intensity of sky continuum is proportional to that of the emission line. Generally, the assumption is
good in dark night within one degree field of view, which is comparable to the field of view of single LAMOST spectrograph.
In the moonlit night, the moonlight  scattered by the atmosphere produce background gradient, which must be taken into account
in sky subtraction.

The sky brightness at LAMOST site is about V=17 mag/arcsec$^2$ at lunar age about 14 days(\cite{Yao13}), which is close to
full moon. Considering the size of the LAMOST fiber, 3.3" in diameter, the moon light dominate  the sky background with V=14.85 mag.
Meanwhile, the faintest targets observed in the bright night is about 14 mag,  2.2 times the background brightness. The typical flux
for a 10-minutes-exposure  V=14 mag star is about 11000 counts/pix, while the sky background  is about 5000 counts/pix, which means 2\% sky background
residual will lead to 1\%  uncertainty in the object flux. Thus a 4\%  sky background gradient across the spectrograph field will lead to
1\% system sky subtraction bias, which is acceptable when the sky subtraction residual is usually larger than 2\%, as in Fig.\ref{skyresidual}.
The sky brightness gradient   depends strongly on the angular distance between the target field and the moon.
If the  sky background gradient inside single spectrograph is limited to less than 4\%, then we cloud define a secure distance to the moon, beyond which
the current sky subtraction scheme is still suitable.

\begin{figure}[bp]
\centering
\includegraphics[width=0.95\textwidth]{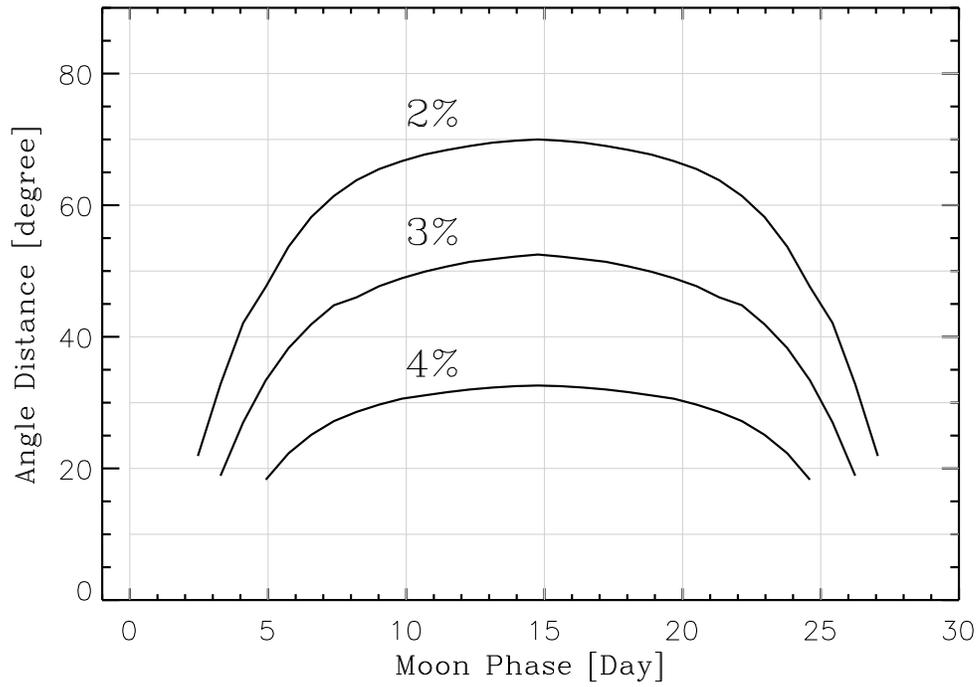}
\caption{The curve of  angle distance between the LAMOST field center and the moon vs the moon phase. The curves represent the angular distance limit
 beyond which the moon light brightness gradient inside individual spectrograph is less than the given value marked above each curve.
The x-axis is the date of moon age, where 0 denotes new moon and 14.7 denotes full moon.
} \label{mdis}
\end{figure}

With a moon night sky brightness model, Yao et al. (2013) calculated the typical sky brightness distribution
for  LAMOST site. According to the model,  the secure angular distances for sky brightness gradient of 2\%, 3\% and 4\% are
derived respectively for different moon phase, as shown in Fig.\ref{mdis}. The angle distance needs to be more
than $33^{\circ}$ to obtain a gradient less than 4\%.
Most LAMOST observations,  both regular survey and test observations at full moon nights, conform to the condition.
A better sky subtraction scheme inducing  the moon light brightness  model will be the future work for the LAMOST 2D pipeline.

\subsection {Optical PSF Variations}
Sky subtraction with the master sky spectrum requires that the shape of the PSF in different fibers are similar, which
is usually unsatisfied due to the imperfect spectrograph optics such as  optical  aberration and distortion, or due to the
 irregular shape of the fiber output end(e.g. due to the poor coupling between the fiber alignment and the slit).
The sky subtraction residual caused by the profile difference in master sky subtraction can not be completely removed by the PCA sky subtraction either.
To further improve the sky subtraction, future works like  careful re-alignment of the fiber and the slit,
and introducing spectra extraction method with 2D de-convolution algorithm (e.g. \cite{Bolton10}, \cite{Li15}) will be necessary.

\section {Conclusion}
\begin{figure}[bp]
\centering
\includegraphics[width=0.95\textwidth]{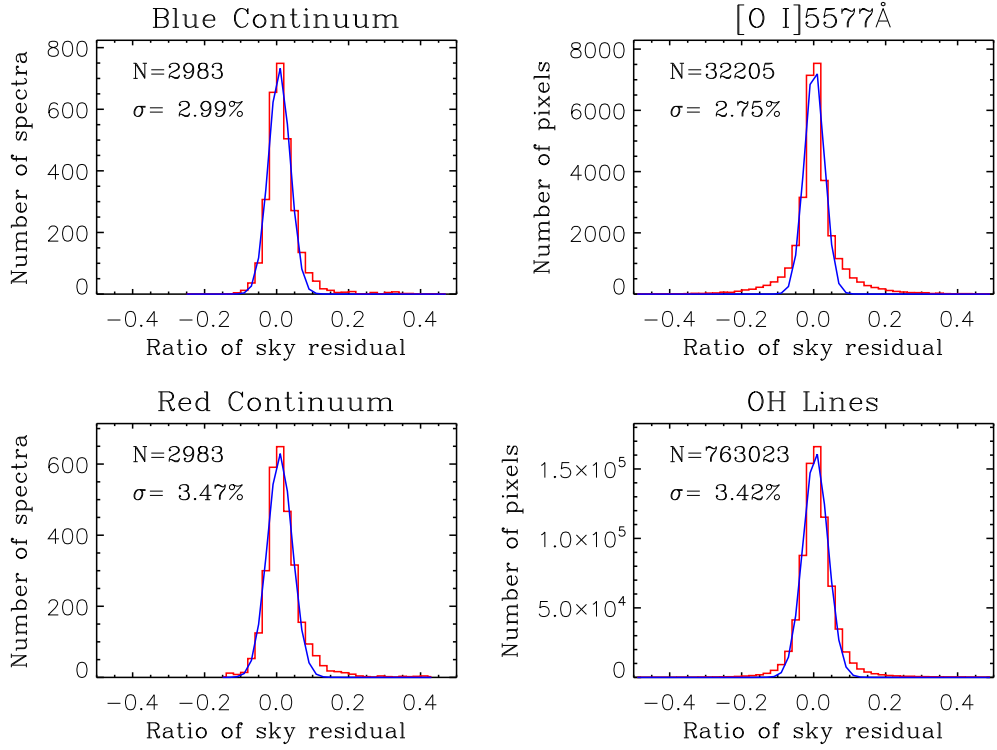}
\caption{Residuals of sky subtracted sky spectra   taken on Sep. 20th, 2016.
The left column is the  residual of the  average continuum (see equation \ref{eq:cave}) in 5470-5560\AA\ and 6000-6200\AA\ for the blue arm and the red arm, respectively.
The right column is the residual of the  individual pixels of sky emission lines.
Each histogram is fitted by a Gaussian function, as shown by the blue curve in each plot.
The total number of pixels and the $\sigma_r$ of the fitting are marked  in the plots.} \label{skyresidualsample}
\end{figure}

Sky subtraction, related to almost every step of the data reduction process, is the most important indicator of the performance of 2D multi-fiber data reduction pipeline.
The key algorithms and the results of sky subtraction in LAMOST 2D pipeline are demonstrated here.
Due to the special characteristic of LAMOST telescope (variable vignetting with telescope pointing) and the  defect in hardware manufacturing and installation
(force added  by fiber positioner to the fibers),  the throughput of LAMOST varies with the telescope pointing and the fiber position, which
lead to the failure of the traditional flat field correction, leaving the sky emission line throughput calibration the current most reliable method.
Sky emission lines are also used to  finetune the wavelength shift about 0.1\AA\   caused by the  instability of the spectrograph.
After the  subtraction by  the finetuned  master sky spectra, the red part of the object spectra are processed by PCA sky subtraction
to further remove the emission line residual of OH band.
Overall, the   wavelength calibration accuracy is about 4.5km/s  according to the RV measurement of the repeat observation  of the same stars.
In dark nights, the median of the sky residual is about 1.5\% and 1.7\% for [OI]5577 and OH lines, respectively. in moonlit nights, the residuals
 rise to  3\% and 3.4\% for [OI]5577 and OH lines respectively, due to  the decrease of the exposure time.
For the sky continuum, the typical relative residuals are about  3\%(Fig.\ref{skyresidualsample}).  As pointed out in Section\ref{scvar},  system bias of sky subtraction  could be limited within 1\%
if the distance between the object and the moon is   larger than $33^{\circ}$.
Our final sky subtracted spectra show smaller  residual in sky emission line region  than SDSS spectra from the analysis of the F-type stars.

\textbf{Aknowledgements}

Thanks for the advices of the reviewers. This work is supported by the National Natural
Science Foundation of China (NSFC) (Grant No. 11503054),  NSFC Key Program (Grant No. 11333004) and the National Key Basic Research
Program of China (Program 973; Grant No. 2014CB845700).

Guoshoujing Telescope (the Large sky Area Multi-Object fiber
Spectroscopic Telescope, LAMOST) is a National Major Scientific Project built by the Chinese
Academy of Sciences. Funding for the project has been provided by the National Development
and Reform Commission. LAMOST is operated and managed by the National Astronomical
Observatories, Chinese Academy of Sciences.

Funding for SDSS-III has been provided by the Alfred P. Sloan Foundation, the Participating Institutions, the National Science Foundation, and the U.S. Department of Energy Office of Science. The SDSS-III web site is http://www.sdss3.org/.

SDSS-III is managed by the Astrophysical Research Consortium for the Participating Institutions of the SDSS-III Collaboration including the University of Arizona, the Brazilian Participation Group, Brookhaven National Laboratory, Carnegie Mellon University, University of Florida, the French Participation Group, the German Participation Group, Harvard University, the Instituto de Astrofisica de Canarias, the Michigan State/Notre Dame/JINA Participation Group, Johns Hopkins University, Lawrence Berkeley National Laboratory, Max Planck Institute for Astrophysics, Max Planck Institute for Extraterrestrial Physics, New Mexico State University, New York University, Ohio State University, Pennsylvania State University, University of Portsmouth, Princeton University, the Spanish Participation Group, University of Tokyo, University of Utah, Vanderbilt University, University of Virginia, University of Washington, and Yale University.

\input{bib.tex}

\clearpage
\begin{sidewaystable}[ht]
\begin{center}
\centering
\caption{Wavelength shift between arc exposures on Jan 1st, 2017. \label{t_waveshift}}
\begin{tabular}{|c|l|l|l|l|l|l|l|l|l|l|l|l|l|l|l|l|}
\hline\hline
spectrograph &  01&   02&   03&    04&    05&    06&    07&    08&    09&    10&    11&    12&    13&    14&    15&    16\\ \hline
Blue 0-1\footnotemark & -0.010 & 0.057 &-0.064 & 0.020&  0.016 & 0.013 & 0.018& -0.104 & 0.046&  0.100& -0.067 & 0.036 & 0.004 & 0.046 & 0.064& -0.032\\
Blue 0-2&-0.033 & 0.081 & -0.089& -0.007 & 0.033 & 0.023 & 0.005& -0.104  &0.044 & 0.152 &-0.074 & 0.046 & 0.013 & 0.032 & 0.114& -0.021\\
Red 0-1&  0.003&  0.116& -0.010& -0.026 & 0.038 &-0.084 & 0.083 &-0.094& -0.017 & 0.097 &-0.108 & 0.031& -0.012 & 0.068 &-0.028& -0.166\\
Red 0-2&  0.009&  0.164 & 0.011& -0.040 & 0.046 &-0.127 & 0.118& -0.023& -0.068 & 0.140& -0.121&  0.043& -0.026 & 0.063 & 0.028& -0.183\\
\hline
\end{tabular}
\footnotetext{Blue and Red denote the blue and red part of the spectrograph; 0-1 is the wavelength shift(in \AA) between the  arc exposure taken at the beginning of the night observation(evening) and that at the middle of the observation(midnight); 0-2 is the difference between the arc at the beginning and that at the end  of the observation(morning).}

\end{center}

\end{sidewaystable}

\clearpage
\begin{table}
\begin{center}
\centering \caption{Sky lines used to measure the image shift and the scale factor.\label{shiftline}}
\begin{tabular}{c|l|l}
\hline\hline
$\lambda$(\AA) & source & Pattern \\
\hline
5577.334 & OI & single \\
6300.304 & OI & single \\
6363.780 & OI & single \\
6863.955 & OI & single \\
6923.220 & OH 7-2 P1 & single \\
7316.282 & OH 8-3 P1 & single \\
7340.885 & OH 8-3 P1 & single \\
7369.366 & OH 8-3 P2 & blend 7369.248 7369.483 \\
7401.858 & OH 8-3 P2 & blend 7401.688 7402.029 \\
7750.640 & OH 9-4 Q1 & single \\
7794.112 & OH 9-4 P1 & single \\
7821.503 & OH 9-4 P1 & single \\
7993.332 & OH 5-1 P1 & single \\
8025.810 & OH 5-1 P2 & blend 8025.668 8025.952 \\
8399.170 & OH 6-2 P1 & single \\
8465.358 & OH 6-2 P2 & blend 8465.208 8465.509 \\
8885.850 & OH 7-3 P1 & single \\
8943.395 & OH 7-3 P2 & single \\
8958.084 & OH 7-3 P2 & blend 8957.922 8958.246 \\
9001.346 & OH 7-3 P2 & blend 9001.115 9001.577 \\
\hline
\end{tabular}
\end{center}
\end{table}

\clearpage

\clearpage

\clearpage

\clearpage

\clearpage

\clearpage

\clearpage

\clearpage

\clearpage

\clearpage

\clearpage

\clearpage

\clearpage

\end{document}

%% file: bib.tex
%\textbf{REFERENCES} \\